\documentclass[aps,prd,twocolumn,showpacs,superscriptaddress,floatfix,nobibnotes,nofootinbib]{revtex4-1}

\usepackage{epsfig}
\usepackage{amsmath}
\usepackage{comment}
\usepackage[labelformat=simple]{subfig}
\usepackage{color}
\usepackage{dcolumn}
\usepackage{multirow}
\usepackage{url}
\usepackage[inline]{enumitem}

\setkeys{Gin}{width=\columnwidth,keepaspectratio}

\usepackage{xspace}

\newcommand{\abseta} {\mid\eta \mid\leq}

\newcommand{\yl}     {y_{\ell}}
\newcommand{\yq}     {q\yl}

\newcommand{\Npartp} {N\left(\yq\right)}
\newcommand{\Npartn} {N\left(-\yq\right)}
\newcommand{\Fpart}  {\mathcal F\left(\yq\right)}
\newcommand{\Apart}  {\mathcal A\left(\yq\right)}
\newcommand{\Apartp} {\mathcal A\left(\yq\right)}
\newcommand{\Apartn} {\mathcal A\left(-\yq\right)}
\newcommand{\Spart}  {\mathcal S\left(\yq\right)}
\newcommand{\Spartp} {\mathcal S\left(\yq\right)}
\newcommand{\Spartn} {\mathcal S\left(-\yq\right)}

\newcommand{\Npartge} {N\left(\yq>0\right)}
\newcommand{\Npartle} {N\left(\yq<0\right)}

\newcommand{\afblep} {A_\mathrm{FB}^{\ell}}

\newcommand{\Apartf} {\afblep\left(\yq\right)}
\newcommand{\Spartf} {S\left(\yq\right)}

\def\urltilde{\kern -.15em\lower .7ex\hbox{\~{}}\kern .04em}
\def\urldot{\kern -.10em.\kern -.10em}
\def\urlhttp{http\kern -.10em\lower -.1ex\hbox{:}\kern -.12em\lower 0ex\hbox{/}\kern -.18em\lower 0ex\hbox{/}}

\def\urltilde{\kern -.15em\lower .7ex\hbox{\~{}}\kern .04em}
\def\urldot{\kern -.10em.\kern -.10em}
\def\urlhttp{http\kern -.10em\lower -.1ex\hbox{:}\kern -.12em\lower 0ex\hbox{/}\kern -.18em\lower 0ex\hbox{/}}

\def\gev {\ifmmode {\rm GeV} \else GeV\fi}
\def\tev {\ifmmode {\rm TeV} \else TeV\fi}
\def\mevc {\ifmmode {\rm MeV}/c \else MeV/$c$\fi}
\def\mevcc {\ifmmode {\rm MeV}/c^2 \else MeV/$c^2$\fi}
\def\gevc {\ifmmode {\rm GeV}/c \else GeV/$c$\fi}
\def\gevcc {\ifmmode {\rm GeV}/c^2 \else GeV/$c^2$\fi}
\def\tevcc {\ifmmode {\rm TeV}/c^2 \else TeV/$c^2$\fi}

\def\ol   {\overline}

\def\vtd  {\ifmmode |V_{td}| \else $|V_{td}|$\fi}
\def\vtb  {\ifmmode |V_{tb}| \else $|V_{tb}|$\fi}
\def\vts  {\ifmmode |V_{ts}| \else $|V_{ts}|$\fi}
\def\vcb  {\ifmmode |V_{cb}| \else $|V_{cb}|$\fi}

\newcommand{\Ds} {\ifmmode D_{\mbox{\sl s}}^{-}
                       \else $D_{\mbox{\sl s}}^{-}$\fi}
\newcommand{\Bs} {\ifmmode B_{\mbox{\sl s}}^{0}
                       \else $B_{\mbox{\sl s}}^{0}$\fi}
\newcommand{\Bsb} {\ifmmode \ol B_{\mbox{\sl s}}^{0}
                       \else $\ol B_{\mbox{\sl s}}^{0}$\fi}
\newcommand{\Bsh} {\ifmmode B_{\mbox{\sl s}}^H
                       \else $B_{\mbox{\sl s}}^H$\fi}
\newcommand{\Bsl} {\ifmmode B_{\mbox{\sl s}}^L
                       \else $B_{\mbox{\sl s}}^L$\fi}
\newcommand{\Dsl} {\ifmmode D_{\mbox{\sl s}}^{-} \ell^+
                       \else $D_{\mbox{\sl s}}^{-} \ell^+$\fi}
\newcommand{\xs} {\ifmmode x_{\mbox{\sl s}}
                       \else $x_{\mbox{\sl s}}$\fi}
\newcommand{\xd} {\ifmmode x_d \else $x_d$\fi}
\newcommand{\lxy} {\ifmmode L_{\rm xy} \else $L_{\rm xy}$\fi}
\newcommand{\dgam} {\ifmmode \Delta\Gamma \else $\Delta\Gamma$\fi}
\newcommand{\dm} {\ifmmode \Delta m \else $\Delta m$\fi}
\newcommand{\ctau} {\ifmmode c\tau \else $c\tau$\fi}
 
\newcommand{\et}{E_T}

\newcommand{\ptran}{\mbox{${p_T}$}}
\newcommand{\etran}{\mbox{${E_T}$}}
\newcommand{\met}{\ensuremath{\protect \raisebox{.3ex}{\(\not\!\)}\et}}

\newcommand{\ppbar}{p\bar{p}}
\newcommand{\qqbar}{q\bar{q}} 
\newcommand{\ttbar}{t\bar{t}}

\newcommand{\afb}{A_\mathrm{FB}^{\Delta \mathrm{y}}}

\newcommand{\dy}{\Delta y}

\newcommand{\ifb}{ {\rm fb}^{-1} }

\begin{document}


\title{\boldmath{Measurement of the leptonic asymmetry in $t\bar{t}$ events produced in $p\bar{p}$ collisions at $\sqrt{s}=1.96~\tev$}}

\affiliation{Institute of Physics, Academia Sinica, Taipei, Taiwan 11529, Republic of China}
\affiliation{Argonne National Laboratory, Argonne, Illinois 60439, USA}
\affiliation{University of Athens, 157 71 Athens, Greece}
\affiliation{Institut de Fisica d'Altes Energies, ICREA, Universitat Autonoma de Barcelona, E-08193, Bellaterra (Barcelona), Spain}
\affiliation{Baylor University, Waco, Texas 76798, USA}
\affiliation{Istituto Nazionale di Fisica Nucleare Bologna, \ensuremath{^{jj}}University of Bologna, I-40127 Bologna, Italy}
\affiliation{University of California, Davis, Davis, California 95616, USA}
\affiliation{University of California, Los Angeles, Los Angeles, California 90024, USA}
\affiliation{Instituto de Fisica de Cantabria, CSIC-University of Cantabria, 39005 Santander, Spain}
\affiliation{Carnegie Mellon University, Pittsburgh, Pennsylvania 15213, USA}
\affiliation{Enrico Fermi Institute, University of Chicago, Chicago, Illinois 60637, USA}
\affiliation{Comenius University, 842 48 Bratislava, Slovakia; Institute of Experimental Physics, 040 01 Kosice, Slovakia}
\affiliation{Joint Institute for Nuclear Research, RU-141980 Dubna, Russia}
\affiliation{Duke University, Durham, North Carolina 27708, USA}
\affiliation{Fermi National Accelerator Laboratory, Batavia, Illinois 60510, USA}
\affiliation{University of Florida, Gainesville, Florida 32611, USA}
\affiliation{Laboratori Nazionali di Frascati, Istituto Nazionale di Fisica Nucleare, I-00044 Frascati, Italy}
\affiliation{University of Geneva, CH-1211 Geneva 4, Switzerland}
\affiliation{Glasgow University, Glasgow G12 8QQ, United Kingdom}
\affiliation{Harvard University, Cambridge, Massachusetts 02138, USA}
\affiliation{Division of High Energy Physics, Department of Physics, University of Helsinki, FIN-00014, Helsinki, Finland; Helsinki Institute of Physics, FIN-00014, Helsinki, Finland}
\affiliation{University of Illinois, Urbana, Illinois 61801, USA}
\affiliation{The Johns Hopkins University, Baltimore, Maryland 21218, USA}
\affiliation{Institut f\"{u}r Experimentelle Kernphysik, Karlsruhe Institute of Technology, D-76131 Karlsruhe, Germany}
\affiliation{Center for High Energy Physics: Kyungpook National University, Daegu 702-701, Korea; Seoul National University, Seoul 151-742, Korea; Sungkyunkwan University, Suwon 440-746, Korea; Korea Institute of Science and Technology Information, Daejeon 305-806, Korea; Chonnam National University, Gwangju 500-757, Korea; Chonbuk National University, Jeonju 561-756, Korea; Ewha Womans University, Seoul, 120-750, Korea}
\affiliation{Ernest Orlando Lawrence Berkeley National Laboratory, Berkeley, California 94720, USA}
\affiliation{University of Liverpool, Liverpool L69 7ZE, United Kingdom}
\affiliation{University College London, London WC1E 6BT, United Kingdom}
\affiliation{Centro de Investigaciones Energeticas Medioambientales y Tecnologicas, E-28040 Madrid, Spain}
\affiliation{Massachusetts Institute of Technology, Cambridge, Massachusetts 02139, USA}
\affiliation{University of Michigan, Ann Arbor, Michigan 48109, USA}
\affiliation{Michigan State University, East Lansing, Michigan 48824, USA}
\affiliation{Institution for Theoretical and Experimental Physics, ITEP, Moscow 117259, Russia}
\affiliation{University of New Mexico, Albuquerque, New Mexico 87131, USA}
\affiliation{The Ohio State University, Columbus, Ohio 43210, USA}
\affiliation{Okayama University, Okayama 700-8530, Japan}
\affiliation{Osaka City University, Osaka 558-8585, Japan}
\affiliation{University of Oxford, Oxford OX1 3RH, United Kingdom}
\affiliation{Istituto Nazionale di Fisica Nucleare, Sezione di Padova, \ensuremath{^{kk}}University of Padova, I-35131 Padova, Italy}
\affiliation{University of Pennsylvania, Philadelphia, Pennsylvania 19104, USA}
\affiliation{Istituto Nazionale di Fisica Nucleare Pisa, \ensuremath{^{ll}}University of Pisa, \ensuremath{^{mm}}University of Siena, \ensuremath{^{nn}}Scuola Normale Superiore, I-56127 Pisa, Italy, \ensuremath{^{oo}}INFN Pavia, I-27100 Pavia, Italy, \ensuremath{^{pp}}University of Pavia, I-27100 Pavia, Italy}
\affiliation{University of Pittsburgh, Pittsburgh, Pennsylvania 15260, USA}
\affiliation{Purdue University, West Lafayette, Indiana 47907, USA}
\affiliation{University of Rochester, Rochester, New York 14627, USA}
\affiliation{The Rockefeller University, New York, New York 10065, USA}
\affiliation{Istituto Nazionale di Fisica Nucleare, Sezione di Roma 1, \ensuremath{^{qq}}Sapienza Universit\`{a} di Roma, I-00185 Roma, Italy}
\affiliation{Mitchell Institute for Fundamental Physics and Astronomy, Texas A\&M University, College Station, Texas 77843, USA}
\affiliation{Istituto Nazionale di Fisica Nucleare Trieste, \ensuremath{^{rr}}Gruppo Collegato di Udine, \ensuremath{^{ss}}University of Udine, I-33100 Udine, Italy, \ensuremath{^{tt}}University of Trieste, I-34127 Trieste, Italy}
\affiliation{University of Tsukuba, Tsukuba, Ibaraki 305, Japan}
\affiliation{Tufts University, Medford, Massachusetts 02155, USA}
\affiliation{University of Virginia, Charlottesville, Virginia 22906, USA}
\affiliation{Waseda University, Tokyo 169, Japan}
\affiliation{Wayne State University, Detroit, Michigan 48201, USA}
\affiliation{University of Wisconsin, Madison, Wisconsin 53706, USA}
\affiliation{Yale University, New Haven, Connecticut 06520, USA}

\author{T.~Aaltonen}
\affiliation{Division of High Energy Physics, Department of Physics, University of Helsinki, FIN-00014, Helsinki, Finland; Helsinki Institute of Physics, FIN-00014, Helsinki, Finland}
\author{S.~Amerio\ensuremath{^{kk}}}
\affiliation{Istituto Nazionale di Fisica Nucleare, Sezione di Padova, \ensuremath{^{kk}}University of Padova, I-35131 Padova, Italy}
\author{D.~Amidei}
\affiliation{University of Michigan, Ann Arbor, Michigan 48109, USA}
\author{A.~Anastassov\ensuremath{^{w}}}
\affiliation{Fermi National Accelerator Laboratory, Batavia, Illinois 60510, USA}
\author{A.~Annovi}
\affiliation{Laboratori Nazionali di Frascati, Istituto Nazionale di Fisica Nucleare, I-00044 Frascati, Italy}
\author{J.~Antos}
\affiliation{Comenius University, 842 48 Bratislava, Slovakia; Institute of Experimental Physics, 040 01 Kosice, Slovakia}
\author{G.~Apollinari}
\affiliation{Fermi National Accelerator Laboratory, Batavia, Illinois 60510, USA}
\author{J.A.~Appel}
\affiliation{Fermi National Accelerator Laboratory, Batavia, Illinois 60510, USA}
\author{T.~Arisawa}
\affiliation{Waseda University, Tokyo 169, Japan}
\author{A.~Artikov}
\affiliation{Joint Institute for Nuclear Research, RU-141980 Dubna, Russia}
% ...

\collaboration{CDF Collaboration}
\altaffiliation[With visitors from]{
\ensuremath{^{a}}University of British Columbia, Vancouver, BC V6T 1Z1, Canada,
\ensuremath{^{b}}Istituto Nazionale di Fisica Nucleare, Sezione di Cagliari, 09042 Monserrato (Cagliari), Italy,
\ensuremath{^{c}}University of California Irvine, Irvine, CA 92697, USA,
\ensuremath{^{d}}Institute of Physics, Academy of Sciences of the Czech Republic, 182~21, Czech Republic,
\ensuremath{^{e}}CERN, CH-1211 Geneva, Switzerland,
\ensuremath{^{f}}Cornell University, Ithaca, NY 14853, USA,
\ensuremath{^{g}}University of Cyprus, Nicosia CY-1678, Cyprus,
\ensuremath{^{h}}Office of Science, U.S. Department of Energy, Washington, DC 20585, USA,
\ensuremath{^{i}}University College Dublin, Dublin 4, Ireland,
\ensuremath{^{j}}ETH, 8092 Z\"{u}rich, Switzerland,
\ensuremath{^{k}}University of Fukui, Fukui City, Fukui Prefecture, Japan 910-0017,
\ensuremath{^{l}}Universidad Iberoamericana, Lomas de Santa Fe, M\'{e}xico, C.P. 01219, Distrito Federal,
\ensuremath{^{m}}University of Iowa, Iowa City, IA 52242, USA,
\ensuremath{^{n}}Kinki University, Higashi-Osaka City, Japan 577-8502,
\ensuremath{^{o}}Kansas State University, Manhattan, KS 66506, USA,
\ensuremath{^{p}}Brookhaven National Laboratory, Upton, NY 11973, USA,
\ensuremath{^{q}}University of Manchester, Manchester M13 9PL, United Kingdom,
\ensuremath{^{r}}Queen Mary, University of London, London, E1 4NS, United Kingdom,
\ensuremath{^{s}}University of Melbourne, Victoria 3010, Australia,
\ensuremath{^{t}}Muons, Inc., Batavia, IL 60510, USA,
\ensuremath{^{u}}Nagasaki Institute of Applied Science, Nagasaki 851-0193, Japan,
\ensuremath{^{v}}National Research Nuclear University, Moscow 115409, Russia,
\ensuremath{^{w}}Northwestern University, Evanston, IL 60208, USA,
\ensuremath{^{x}}University of Notre Dame, Notre Dame, IN 46556, USA,
\ensuremath{^{y}}Universidad de Oviedo, E-33007 Oviedo, Spain,
\ensuremath{^{z}}CNRS-IN2P3, Paris, F-75205 France,
\ensuremath{^{aa}}Universidad Tecnica Federico Santa Maria, 110v Valparaiso, Chile,
\ensuremath{^{bb}}The University of Jordan, Amman 11942, Jordan,
\ensuremath{^{cc}}Universite catholique de Louvain, 1348 Louvain-La-Neuve, Belgium,
\ensuremath{^{dd}}University of Z\"{u}rich, 8006 Z\"{u}rich, Switzerland,
\ensuremath{^{ee}}Massachusetts General Hospital, Boston, MA 02114 USA,
\ensuremath{^{ff}}Harvard Medical School, Boston, MA 02114 USA,
\ensuremath{^{gg}}Hampton University, Hampton, VA 23668, USA,
\ensuremath{^{hh}}Los Alamos National Laboratory, Los Alamos, NM 87544, USA,
\ensuremath{^{ii}}Universit\`{a} degli Studi di Napoli Federico I, I-80138 Napoli, Italy
}
\noaffiliation

\date{\today}

\begin{abstract}
We measure the asymmetry in the charge-weighted rapidity $\yq$ of the lepton in semileptonic $\ttbar$ decays recorded with the CDF II detector using the full Tevatron Run II sample, corresponding to an integrated luminosity of $9.4 ~\ifb$.  A parametrization of the asymmetry as a function of $\yq$ is used to correct for the finite acceptance of the detector and recover the production-level asymmetry.  The result of $\afblep = 0.094^{+0.032}_{-0.029}$ is to be compared to the standard model next-to-leading-order prediction of $\afblep=0.038\pm0.003$.

\end{abstract}

\maketitle

\section{Introduction\label{Intro}}

The CDF and D0 experiments have reported a large forward-backward asymmetry in top-quark pair-production in $\ppbar$ collisions at $\sqrt{s} = 1.96$~\tev~\cite{Aaltonen:2011kc,Abazov:2011rq}. The asymmetry is measured in the $\ttbar$ rapidity difference $\dy$, reconstructed in event topologies involving final states with a single charged lepton and hadronic jets ($\ell$+jets) or two charged leptons and hadronic jets (dilepton).  The most recent CDF measurement finds $\afb = 0.164 \pm 0.045$, compared to the prediction of $\afb = 0.066 \pm 0.020$, which includes both electroweak and next-to-leading-order (NLO) QCD effects~\cite{Aaltonen:2012it}.  D0 measures $\afb = 0.196 \pm 0.065$~\cite{Abazov:2011rq}.  Measurements in $pp$ collisions of the top-quark charge asymmetry $A_C$, an observable that is distinct from $\afb$ but correlated with it, have found higher consistency with the standard model (SM)~\cite{ ATLAS:2012an, Chatrchyan:2012cxa}.  However, any observable effect at the LHC is expected to be small, and the nature of the relationship between $\afb$ and $A_C$ is model-dependent~\cite{AguilarSaavedra:2012va, PhysRevD.86.054022, PhysRevD.86.114034, PhysRevD.86.094040, Berger:2012tj}. 

These measurements rely on the reconstruction of the top-quark direction in complex final states with leptons, jets, and an azimuthal imbalance in the total transverse momentum in the event (missing energy).  A significant and calculable correlation exists between the direction of a top-quark and its decay products, so that an asymmetry in the parent top-quark direction will induce an asymmetry in the decay products. It is therefore interesting to investigate if an asymmetry in a decay-product direction, which is accessible through simpler analysis, supports the effect previously seen through more complex top-quark decay reconstruction, possibly providing further information on the asymmetry itself.

Amongst the possible top-quark decay products in $\ell$+jets, the lepton is uniquely suited for the measurement of such an asymmetry.  The lepton direction is measured with high precision, and the good charge determination unambiguously identifies whether the lepton's parent quark was a top or antitop.  Furthermore, the leptonic asymmetry $\afblep$ depends on both the top-quark pair asymmetry and the top-quark polarization. Several authors have noted that explanations of the Tevatron asymmetry that include polarized top quarks could lead to measurable changes in the leptonic asymmetry, while leaving unchanged the top-quark pair forward-backward asymmetry~\cite{Falkowski:2012cu, Berger:2012tj, Berger:2013et}.  Such theories predict very different values for $\afblep$, while having similar top-quark asymmetries.  The asymmetry of the lepton is therefore an observable that is usefully correlated with $\afb$, but may also contain additional information on the nature of the top-quark pair asymmetry.

The lepton asymmetry is defined using its electric charge $q$ and rapidity in the lab frame,
\begin{equation} \label{eq:rap}
   \yl = \frac{1}{2} \ln \left( \frac{E + p_z}{E - p_z} \right),
\end{equation}

\noindent where $E$ is the total energy of the lepton and $p_z$ its momentum in the direction of the proton beam.  If charge-parity symmetry ({\it CP}) is conserved, then for leptons of opposite charge, the effects on the lepton rapidity from both the top-quark asymmetry and a possible polarization are equal in magnitude but opposite in sign.  We define a charge-weighted lepton asymmetry,

\begin{equation}
 \afblep = \frac{N\left(\yq>0\right) - N\left(\yq<0\right)}
                {N\left(\yq>0\right) + N\left(\yq<0\right)}.
\end{equation}

This asymmetry has been calculated to NLO, including both QCD and electroweak effects, to be $\afblep = 0.038\pm0.003$~\cite{Bernreuther:2012sx}. The D0 collaboration has measured the asymmetry using a sample corresponding to $5.4~\ifb$ in both $\ell$+jets and dilepton decays, and finds a combined lepton asymmetry of $0.111\pm 0.036$~\cite{Abazov:2011rq,Abazov:2012bfa}.

We report on a measurement of the lepton asymmetry $\afblep$, using the full Tevatron Run II data set of $\sqrt{s}=1.96~\tev$ proton-antiproton collisions as recorded by the CDF II detector~\cite{Acosta:2004yw} at the Fermilab Tevatron and corresponding to an integrated luminosity of $9.4 ~\ifb$.  This measurement is performed in a superset of the $\ell$+jets sample used in the measurement of $\afb$ in Ref.~\cite{Aaltonen:2012it}. That measurement employed a full $\ttbar$ reconstruction and corrected the observed asymmetry to determine the asymmetry at production ({\it production-level}) using a procedure based on singular-value decomposition ~\cite{svd}.  Here we determine $\afblep$ using only the charged lepton.  We examine the expected distributions of $\afblep$ in a number of simulated data samples representing the SM prediction, as well as some non-standard models with new $\ttbar$ production mechanisms and top-quark polarizations (Sec.~\ref{IntroNPModels}).  We show that the charge-weighted lepton rapidity $\yq$ can be separated into a symmetric part $\Spart$, which is largely insensitive to the physics model, and an antisymmetric part $\Apart$ that encapsulates the variation from one model to the next (Sec.~\ref{RapidityDecomp}).  We show that $\Apart$ may be approximated by a simple mathematical form.  We fit this functional dependence in the measured $\Apart$ distribution, and use this in conjunction with the symmetric part taken from simulated models to extract the inclusive production-level $\afblep$ (Sec.~\ref{InclResults}).

\section{Physics Models and Expected Asymmetry\label{IntroNPModels}}

The analysis techniques are designed and validated using model data sets created with Monte Carlo event generators.  Leading order (LO) event generators are configured to use the {\sc cteq6.1L} set of parton-distribution functions (PDFs), while NLO event generators use {\sc cteq6.1M}.  The generated partons are processed by the {\sc pythia}~\cite{Sjostrand:2006za} parton-showering and hadronization algorithms into final-state particles, which are then processed with a full simulation of the CDF II detector.  The effects of the parton shower and hadronization are included in all of the production-level results.

At LO the expected top-quark asymmetry is zero. The NLO QCD asymmetry arises in the interference of $\qqbar$ annihilation diagrams that have opposite behavior under charge conjugation at LO and NLO.  The $gg$ initial state does not contribute to the asymmetry, but does dilute the average value.  To study the SM at the leading order (LO), we use events generated by {\sc alpgen}~\cite{Mangano:2002ea}.  The benchmark for SM $\ttbar$ production at NLO is the {\sc powheg}~\cite{Frixione:2007nw} generator, which includes NLO QCD but not electroweak effects.  We treat {\sc powheg} as the nominal model for all variables of interest except for $\afblep$ itself, for which the calculation of Ref.~\cite{Bernreuther:2012sx}, which explicitly includes electroweak interference effects, is better suited.

\begin{table*}
\caption{Production-level Monte Carlo asymmetries and polarizations.  The uncertainty on the final digit is shown in parentheses.}\label{tab:GenAFB}
\begin{tabular}{lccl}
\hline
\hline
Model                  & $\afb$                 & $\afblep$              &                          \\
\hline
NLO QCD ({\sc powheg}) & $+0.052\left(0\right)$ & $+0.024\left(0\right)$ &                          \\
LO QCD ({\sc alpgen})  & $-0.000\left(1\right)$ & $+0.003\left(1\right)$ &                          \\
Octet A                & $+0.156\left(1\right)$ & $+0.070\left(2\right)$ & LO unpolarized axigluon  \\
Octet L                & $+0.121\left(1\right)$ & $-0.062\left(1\right)$ & LO left-handed axigluon  \\
Octet R                & $+0.114\left(2\right)$ & $+0.149\left(2\right)$ & LO right-handed axigluon \\
\hline
\hline
\end{tabular}

\end{table*}

To study larger asymmetries, we use the {\sc madgraph}~\cite{Alwall:2007st} generator to produce three models containing heavy color-octet partners to the gluon.  The  gluon partners can have axial couplings to the quarks (thus ``axigluons''), and interfere with gluons to produce a top-quark production asymmetry. These models are tuned to explore the lepton asymmetry in three different top-quark polarization scenarios, while maintaining an inclusive $\dy$ asymmetry compatible with Tevatron measurements.  The three models include the cases of new physics contributions with axial-vector couplings between the axigluon and quarks (Octet A), left-handed couplings (Octet L), and right-handed couplings (Octet R).  Octet A includes a massive ($M_A = 2.0~\tevcc$) axigluon~\cite{Aaltonen:2011kc}.  Octet L and Octet R are the models of Ref.~\cite{Falkowski:2012cu}. Both include axigluons of mass $M_A = 200~\gevcc$ and decay width $\Gamma_A = 50~\gevcc$.  The large width is proposed by the authors as a means to evade dijet resonance searches.  However, the importance of these samples in this work is in the validation of the analysis procedures for any polarization and asymmetry, independent of any limits on these particular models.

The lepton asymmetries in these three cases are shown in Table~\ref{tab:GenAFB} along with the SM LO ({\sc alpgen}) and NLO ({\sc powheg}) estimates. The distribution in the charge-weighted lepton rapidity $\yq$ is shown in Fig.~\ref{fig:GenQYLep}.  The lepton asymmetry in Octet A results only from the SM kinematic correlation with $\afb$.  In the right-handed Octet R, top-quark pairs are produced with the spin of both the top and antitop quarks preferentially aligned in the direction of the initiating light quark.  The decay of a top (antitop) quark with such a polarization favors the production of leptons with $\yl>0$ ($\yl<0$), producing an additional positive contribution to the asymmetry of $\yq$.  In Octet L, the negative contribution of the left-handed polarization overcomes the effect of a positive $\afb$ and results in a negative $\afblep$.

\begin{figure}[t]
\includegraphics[clip]{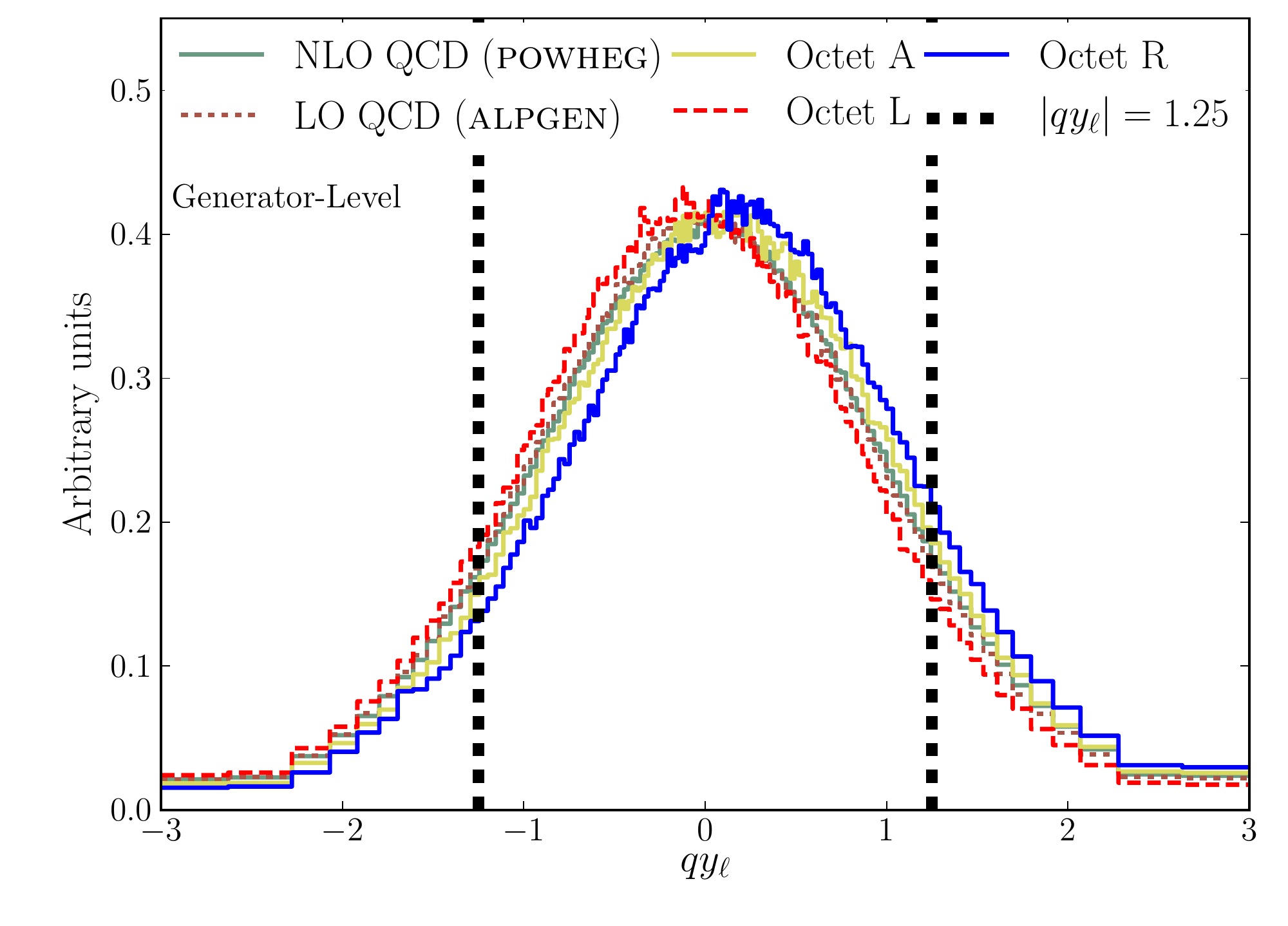}
\caption{{\small The distribution of simulated $\ttbar$ events vs. $\yq$ at the production-level for several models.  The vertical lines at $\left|\yq\right|=1.25$ indicate the limits of the lepton acceptance.}\label{fig:GenQYLep}}
\end{figure}

In light of the correlation between $\afblep$ and $\afb$, it is desirable to have some expectation for $\afblep$ given the measured value of $\afb$.  In general the relationship is model-dependent.  However, in the case where the only substantial deviation from the SM predictions is $\afb$, with no polarization and top-quark decays as described by the SM, an estimate is straightforward.  This includes the cases of either the unpolarized axigluon model discussed above or purely-SM proposals in which unexpectedly large QCD corrections result in an enhanced $\afb$.

One estimate is provided by Octet A, with a top-quark asymmetry of $0.156$, which compares well to the CDF measurement of $0.164\pm 0.047$~\cite{Aaltonen:2012it}.  Octet A predicts no top-quark polarization, so $\afblep$ is entirely due to the kinematic correlation with $\dy$.  The predicted asymmetry of Octet A, $\afblep=0.070$, therefore provides a possible expectation for the data.

A second estimate is derived from the predicted ratio $\afblep/\afb$ in conjunction with the observed value of $\afb$.  When the top quark is unpolarized and decays as the SM top quark, this ratio is fixed.  It may be derived from several sources to confirm the sensibility of this procedure.  The ratio from {\sc powheg} is $0.46$.  The calculation of Ref.~\cite{Bernreuther:2012sx}, which includes predictions for $\afb$ as well as $\afblep$, yields a ratio of $0.43$.  Octet A, which has much larger asymmetries than either of these, has a ratio of $0.45$.  The similarity of these values suggests that a simple ratio is sufficient to capture the kinematic correlation between the two asymmetries.  Given the value $\afb=0.164$ measured by CDF, the expected asymmetry of the lepton calculated with the {\sc powheg} ratio is $0.076$. The concordance of Octet A and ratio-based estimates suggests that a possible expectation for $\afblep$, given no top-quark polarization and the value of $\afb$ measured by CDF, is in the range of $0.070$--$0.076$.

\section{Selection and Background Modeling\label{SelectionAndBG}}

\subsection{Event Selection and Sample Composition}\label{InclEventSelection}

The CDF II detector is a general purpose, azimuthally and forward-backward symmetric magnetic spectrometer with calorimeters and muon detectors~\cite{Acosta:2004yw}. Charged particle trajectories (tracks) are reconstructed with a silicon-microstrip detector and a large open-cell drift chamber in a 1.4 T solenoidal magnetic field.  Projective-tower-geometry electromagnetic and hadronic calorimeters located beyond the solenoid provide electron, jet, and missing energy reconstruction~\cite{coords}. Beyond the calorimeter are multilayer proportional chambers that provide muon detection and identification in the region $\abseta 1.0$.  Electrons are identified by matching isolated charged-particle tracks to clusters of energy deposited in the electromagnetic calorimeter.  We use a cylindrical coordinate system with the origin at the center of the detector and the $z$-axis along the direction of the proton beam~\cite{coords}.

We use the full CDF Run II data set, corresponding to an integrated luminosity of $9.4~\ifb$.  Online, an electron and muon event-selection system (triggers) select candidates with a charged lepton and jets in the final state.  Lepton+jets candidate events are selected from high-$\ptran$ electron or muon triggers.  Additionally, we include events triggered by large missing $E_{T}$ in which a high-$\ptran$, isolated muon is identified through offline reconstruction.  Jets are reconstructed using a cone algorithm~\cite{PhysRevLett.68.1104} with cone radius $R\equiv\sqrt{\left(\Delta\eta\right)^2 + \left(\Delta\phi\right)^2}=0.4$.  The {\sc secvtx} algorithm~\cite{secvtx} is used to identify jets that likely originated from bottom quarks by searching for displaced decay vertices within the jet cones ({\it $b$-tag}) .

After offline event reconstruction, we require that each candidate event contains exactly one electron or muon with $\ptran > 20$~\gevc ~and $\left|\eta\right| < 1.25$.  The maximum pseudorapidity of the lepton is determined by the limited central-tracking acceptance of the CDF II detector.  Extrapolation into the unmeasured region of high lepton pseudorapidity motivates much of the approach of this analysis.  We require $\met > 20$~\gev, consistent with the presence of an undetected neutrino.  We require four or more energetic jets with $\left|\eta\right| < 2.0$.  At least three must have $\etran > 20$~\gev, and the remaining jet(s) must have $\etran > 12$~\gev.  One or more jets with $\etran > 20$~\gev ~must be $b$-tagged.  Finally, we require that $H_T$, the scalar sum of the missing $E_{T}$ plus the transverse energy of the lepton and all jets, be at least $220$~\gev.  This selection extends that of Ref.~\cite{Aaltonen:2012it}, which required that all four jets have $\etran > 20$~\gev.

Models for the non-$\ttbar$ backgrounds are well understood in precision cross-section measurements such as Ref.~\cite{Aaltonen:2011zma}, and provide accurate measures of both the normalizations and shapes of the non-$\ttbar$ processes.  The {\sc alpgen} generator is used to model $W$+heavy flavor ($W$+HF) and $W$+light flavor ($W$+LF) backgrounds.  Small electroweak backgrounds ($Z$+jets, single top quark, and diboson production) are modeled using {\sc pythia}.  The remaining background consists of events in which jets are produced without an associated on-shell gauge boson, and a track is incorrectly identified as an isolated high-$p_{T}$ lepton.  This ``Non-$W$/$Z$'' background is not amenable to simulation.  It is instead modeled using a data-driven sideband taken from events that fail the lepton selection requirements.  The final sample for analysis consists of $3864$ events.  The predicted composition is shown in Table~\ref{tab:method2}; the total background contribution is estimated to be $1026 \pm 210$ events. Further details on the sample, selection, and backgrounds can be found in Ref.~\cite{Aaltonen:2012it}.

\begin{table}
\caption{Estimated sample composition.  The $\ttbar$ yield assumes a production cross-section of $7.4$~pb.}\label{tab:method2}
\begin{tabular}{lrcr}
\hline
\hline
Process           & \multicolumn{3}{c}{Prediction}   \\
\hline
$W$+HF           & $ 481$ & $\pm$ & $ 178$ \\
$W$+LF           & $ 201$ & $\pm$ & $  72$ \\
$Z$+jets         & $  34$ & $\pm$ & $   5$ \\
Single top       & $  67$ & $\pm$ & $   6$ \\
Diboson          & $  36$ & $\pm$ & $   4$ \\
Non-$W$/$Z$      & $ 207$ & $\pm$ & $  86$ \\
\hline
All backgrounds  & $1026$ & $\pm$ & $ 210$ \\
$\ttbar$ (7.4~pb) & $2750$ & $\pm$ & $ 426$ \\
\hline
Total prediction & $3776$ & $\pm$ & $ 476$ \\
Observed         & $3864$ &      &  \\
\hline
\hline
\end{tabular}

\end{table}

\subsection{Treatment of Non-$\ttbar$ Backgrounds\label{BackgroundAsym}}

Non-$\ttbar$ background processes are expected to contribute a non-zero asymmetry to the sample.  This is accounted for by subtracting the expected backgrounds bin-by-bin from the observed distribution of $\yq$.  The largest contribution is from $W$+jets, which is both the dominant background and inherently asymmetric.  The asymmetry in $W$ production arises from various sources.  A negative asymmetry is contributed by the electroweak V-A coupling, but a positive asymmetry arises from $u$-type quarks carrying more momentum on average than $d$-type quarks.  When the $W$ boson is produced in conjunction with jets, a similar imbalance in the momenta of quarks and gluons in $qg$-initiated processes provides an additional positive contribution.

Before performing the background subtraction, we ensure that the background and its asymmetry are properly modeled. This is accomplished by examining events that otherwise meet the criteria of Sec.~\ref{InclEventSelection}, but have exactly zero $b$-tagged jets.  This {\it zero-tag} selection yields a sample that is independent from the signal-region sample, while having very similar kinematic properties, and provides a control region which is substantially enriched in background processes.

\begin{figure}[!h]
\includegraphics[clip]{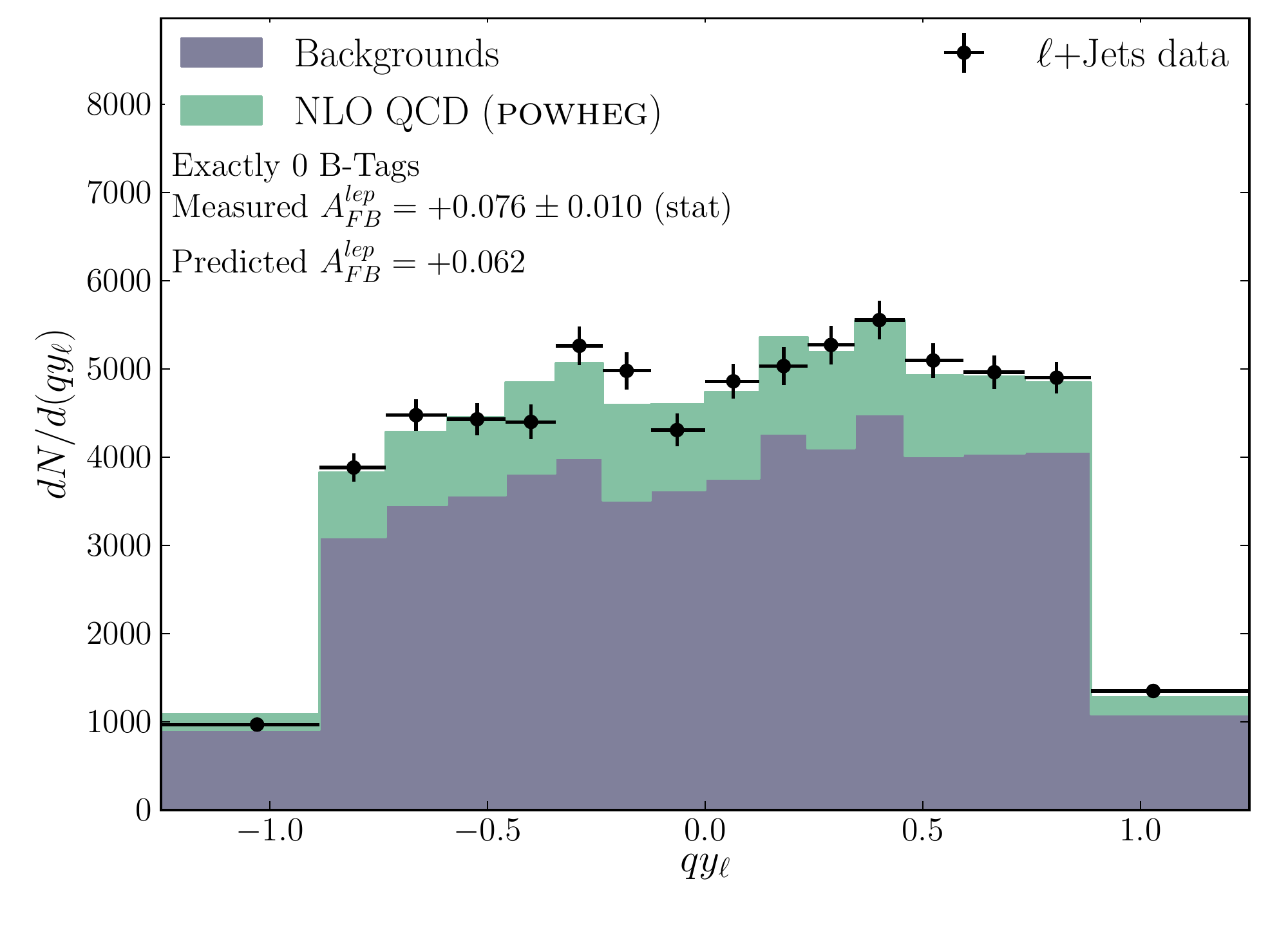}
\caption{{\small The distribution of events vs. $\yq$ in the zero-tag control sample.  Black markers indicate the data.  The filled region represents the prediction of $\ttbar$ (light-colored fill) and backgrounds (dark).} \label{fig:ModelValidationLeptonAntitag}}
\end{figure}

\begin{table}[!htb]
\caption{Comparison of the predicted and measured asymmetries in the zero-tag control sample. ``Signal + backgrounds'' is the predicted asymmetry when the $\afblep$ of the $\ttbar$ component is fixed to $0.070$.}\label{tab:antitag_asyms}
\begin{tabular}{lrcl}\hline
\hline
                           & \multicolumn{3}{c}{Asymmetry} \\
\hline
NLO SM               & $0.017 $ && \\
Backgrounds                & $0.074 $ && \\
\hline
NLO SM + backgrounds & $0.062 $ && \\
\hline
Signal + backgrounds      & $0.073 $ && \\
Data                   & $0.076 $ & $\pm$ & $0.010$ \\
\hline
\hline
\end{tabular}

\end{table}

Figure~\ref{fig:ModelValidationLeptonAntitag} shows the distribution of events as a function of $\yq$ in the zero-tag control sample.  The asymmetries of $\ttbar$ and backgrounds in the control sample are summarized in Table~\ref{tab:antitag_asyms}.  The $\afblep$ predicted by the expected $\ttbar$ and backgrounds is $0.062$, while the asymmetry observed in the data is $0.076 \pm 0.010$, already an acceptable level of agreement.  However, approximately $20\%$ of the control sample consists of top-quark pairs.  As a consistency check, we anticipate the measurement of the background-subtracted asymmetry in the tagged region ($\afblep=0.070$; see Sec.~\ref{InclAsym}).  If the $\ttbar$ component is assumed to have this asymmetry, the predicted $\afblep$ in the control sample becomes $0.073$, in excellent agreement with the measured value, suggesting that this background model is robust.

\section{Methodology\label{MethodDecomp}}

The raw asymmetry includes contributions from non-$\ttbar$ backgrounds and is further distorted by limited detector acceptance.  Both of these effects must be corrected in order to determine the asymmetry at production.  Contributions from the backgrounds are removed using a bin-by-bin subtraction procedure (Sec.~\ref{BackgroundAsym}).  Acceptance corrections must accommodate the steep decline of the acceptance in $\yl$ (Figs.~\ref{fig:GenQYLep} and ~\ref{fig:GenDecomposition}) due to the geometry of the detector. The approximately $20\%$ of events that fall outside the detector's acceptance are also predicted to have the largest forward-backward asymmetry.  The recovery of the production-level inclusive $\afblep$ must necessarily rely on an extrapolation into this unmeasured region.

\subsection{Rapidity Decomposition\label{RapidityDecomp}}

\begin{figure*}
  \subfloat[][]{
    \includegraphics[clip]{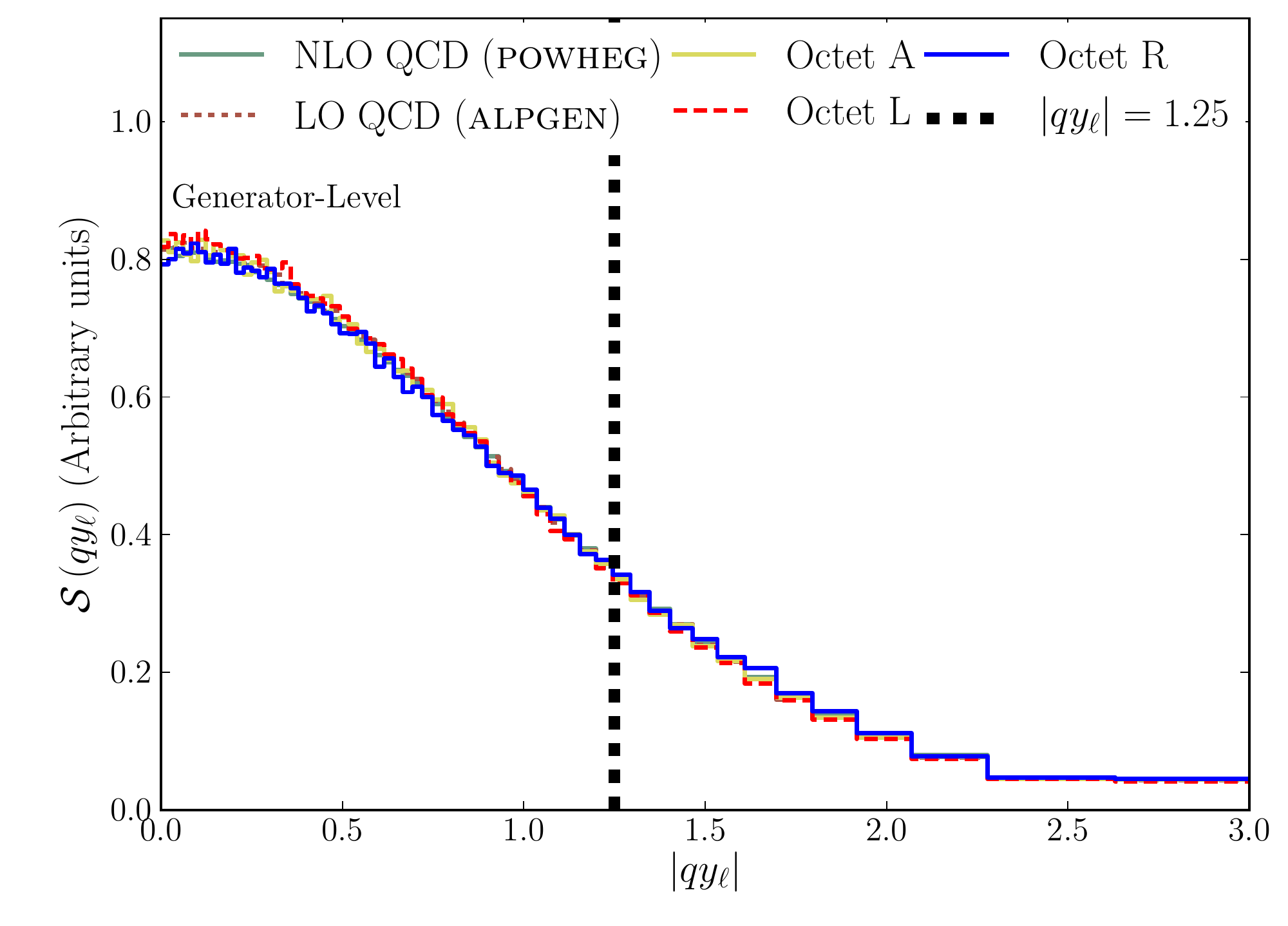}
    \label{fig:GenDecompositionS}
  }
  \subfloat[][]{
    \includegraphics[clip]{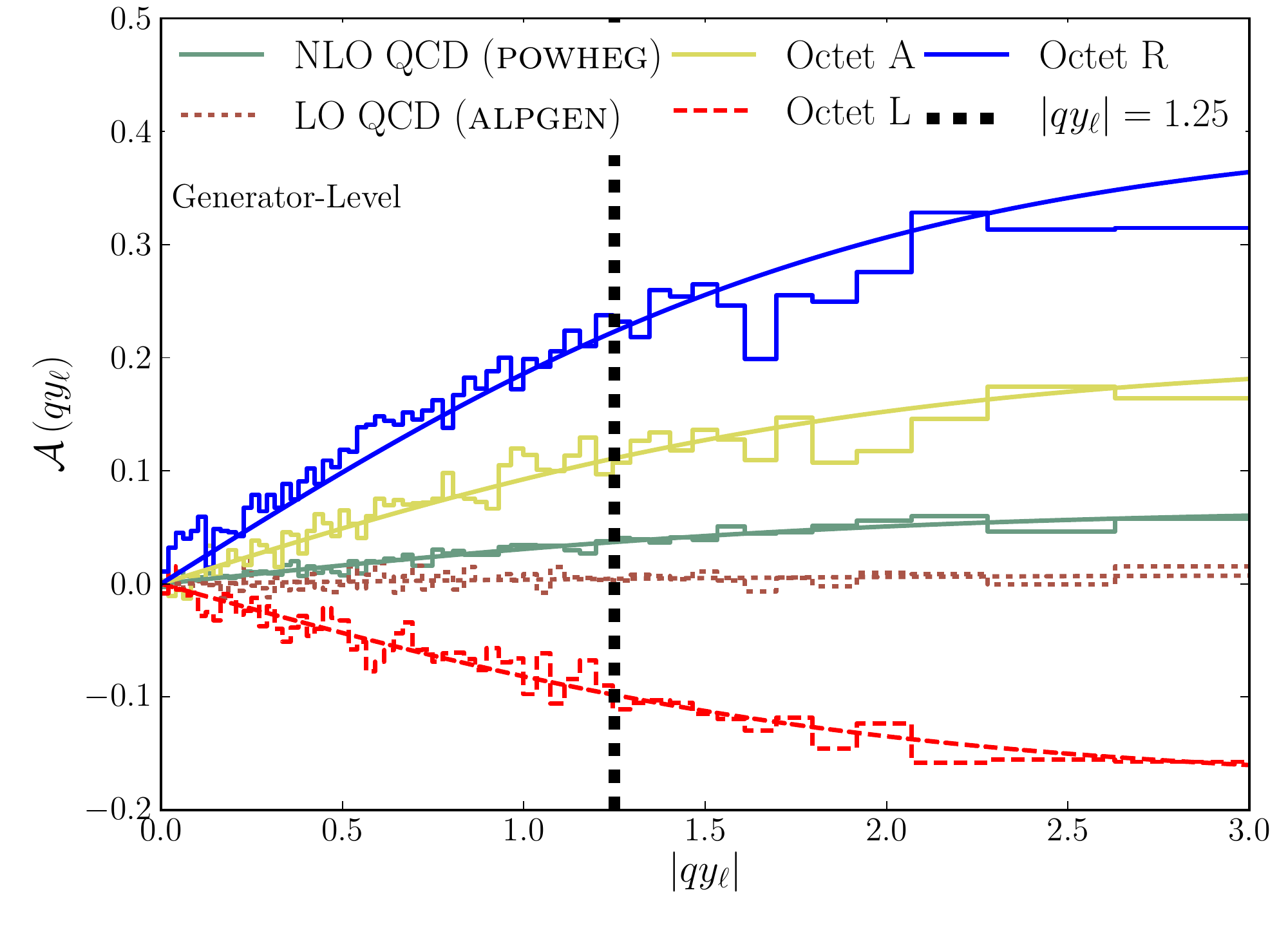}
    \label{fig:GenDecompositionA}
  }
\caption{ The symmetric part \protect\subref*{fig:GenDecompositionS} and asymmetry \protect\subref*{fig:GenDecompositionA} of the production-level distribution of $\yq$ for the discussed models.  Shown also are the best fits to Eq.~(\ref{eqn:APartFit}). The vertical lines at $\left|\yq\right|=1.25$ indicate the limits of the lepton acceptance.\label{fig:GenDecomposition}}
\end{figure*}

The extrapolation relies on a separation of the signed rapidity distribution $\Npartp$ into its symmetric and antisymmetric parts $\Spart$ and $\Apart$, defined as

\begin{subequations}
\begin{align}
\Spart & = \frac {\Npartp+\Npartn}
                 {2}               \\
\Apart & = \frac {\Npartp-\Npartn}
                 {\Npartp+\Npartn}
,\end{align}
\end{subequations}

\noindent in the range $\yq \geq 0$.  Note that $\Apart$ measures the differential dependence of the asymmetry $\Apartf$~\cite{conventions}. The functions $\Spart$ and $\Apart$ may be inverted to recover the original distribution:

\begin{equation}
\Npartp = \begin{cases}
    \Spartp\times \left[ 1 + \Apartp \right] & \yq > 0 \\
    \Spartn\times \left[ 1 - \Apartn \right] & \yq < 0
  .\end{cases}
\end{equation}

\noindent This in turn may be integrated to recover the total number of forward or backward events

\begin{subequations}
\begin{align}
  \Npartge &= \int\limits_{0}^{\infty}d\yq\left\{\Spart\times\left[1 + \Apart\right]\right\} \\
  \Npartle &= \int\limits_{0}^{\infty}d\yq\left\{\Spart\times\left[1 - \Apart\right]\right\}
,\end{align}
\end{subequations}

\noindent which then yields the inclusive asymmetry, written in terms of $\Spart$ and $\Apart$

\begin{subequations}\label{eqn:AfbIntegral}
\begin{align}
 \afblep &= \frac {\Npartge - \Npartle}
                  {\Npartge + \Npartle}\label{eqn:AfbIntegralA} \\
         &= \frac {\int\limits_{0}^{\infty}d\yq \left[\Apart\times\Spart\right]}
                  {\int\limits_{0}^{\infty}d\yq \Spart}\label{eqn:AfbIntegralB}
.\end{align}
\end{subequations}
\noindent

\subsection{Extrapolation Procedure\label{MethodProcedure}}

Figure~\ref{fig:GenDecomposition} shows the shape of the symmetric \subref*{fig:GenDecompositionS} and asymmetric \subref*{fig:GenDecompositionA} parts in the Monte Carlo models.  The shape of $\Spart$ is very similar across models, suggesting little or no dependence on either the top-quark production asymmetry or polarization, while $\Apart$ captures the variation between models.

The form of this decomposition suggests a strategy for extrapolating the asymmetry into the unmeasured region: if $\Apart$ can be parametrized such that its full dependence may be extracted from the measured asymmetry in the accepted region, then the integral of Eq.~(\ref{eqn:AfbIntegralB}) can be used to recover the production-level asymmetry by integrating the measured dependence of $\Apart$ against the predicted production-level $\Spart$ from simulation.

The predictions of $\Apart$ of the models shown in Fig.~\ref{fig:GenDecomposition} are described adequately by the function

\begin{equation}\label{eqn:APartFit}
\Fpart = a \tanh\left(\frac{\yq}{2}\right)
\end{equation}

\noindent and the best-fit curves for this functional form are shown overlaid on the models in Fig.~\ref{fig:GenDecompositionA}. This empirical parametrization is not expected to be completely model-independent.  However, it reproduces the dependence of the asymmetry on $\yq$ for the models discussed here.  In particular, the dependence predicted by the {\sc powheg} generator is accurately described ($\chi^{2}/{\rm ndf}=158/119$), and it is therefore reasonable to expect this functional form to be reliable for any model with kinematic properties sufficiently resembling the SM.  In the next section we show that this choice of parametrization is able to accurately recover the correct production-level asymmetry for all of the considered models.

The procedure to extract the production-level $\afblep$ from data is then the following:
\begin{enumerate*}[label*=(\arabic*)]
\item subtract the expected background contribution in each bin of $\yq$;
\item using acceptances derived from {\sc powheg}, perform bin-by-bin acceptance corrections on the background-subtracted data;
\item fit the acceptance-corrected $\Apartf$ to the functional form $\Fpart$ (Eq.~(\ref{eqn:APartFit}));
\item integrate $\Fpart$ with the $\Spart$ determined in simulation to recover the inclusive $\afblep$.
\end{enumerate*}

The binning of $\yq$ in the data is chosen so that {\sc powheg}'s predicted $\Spart$ equally populates each bin.  The predicted bin centers are calculated as a weighted average of $\yq$ in each bin according to {\sc powheg}.  The fit to $\Apart$ uses this binning and $\Fpart$ evaluated at the predicted bin centers.  Once the fit parameter $a$ of Eq.~(\ref{eqn:APartFit}) is obtained from the background-subtracted data using this binning, the integration of Eq.~(\ref{eqn:AfbIntegral}) is carried out using the 120-bin production-level $\Spart$ values from {\sc powheg}.

\subsection{Validation\label{MethodValidation}}

The efficacy of the correction procedure is tested for each of the models described in Sec.~\ref{IntroNPModels}, using $10~000$ simulated experiments with the $\ttbar$ event yield as in the data.  In each experiment the number of events in each $\yq$ bin is fluctuated according to Poisson statistics, and the acceptance correction and extrapolation procedure is performed to yield a corrected asymmetry that is compared to the known production-level value.

The mean values of the asymmetries in the $10~000$ simulated experiments for each model are shown in Table~\ref{tab:InclAfbTest}.  The extrapolation procedure is successful at recovering the true asymmetry while introducing only minimal model-dependent biases: Absolute deviations of the mean extrapolated result from the true asymmetry are below $0.01$. Note, in particular, that the procedure yields the vanishing asymmetry in the LO standard model, and that biases with the NLO standard model and Octet A (which has an $\afblep$ value similar to that observed in the data) are very small.

\begin{table}[h]
\caption{True asymmetries as generated in simulation compared to mean extrapolated results for $10~000$ simulated experiments with the yield of the $\ttbar$ component as in the data.  The uncertainties on the mean extrapolated results are negiligible compared to the mean values.}\label{tab:InclAfbTest}
\begin{tabular}{lcc}
\hline
\hline
Signal model           & True $\afblep$ & Extrapolated $\afblep$\\
\hline
NLO QCD ({\sc powheg}) & $+0.024$       & $+0.026$ \\
LO SM ({\sc alpgen})   & $+0.003$       & $-0.004$ \\
Octet A                & $+0.070$       & $+0.070$ \\
Octet L                & $-0.062$       & $-0.062$ \\
Octet R                & $+0.149$       & $+0.155$ \\
\hline\hline
\end{tabular}

\end{table}

\section{Measurement of $\afblep$\label{InclResults}}

\subsection{Central Value\label{InclAsym}}

\begin{figure}
\includegraphics[clip]{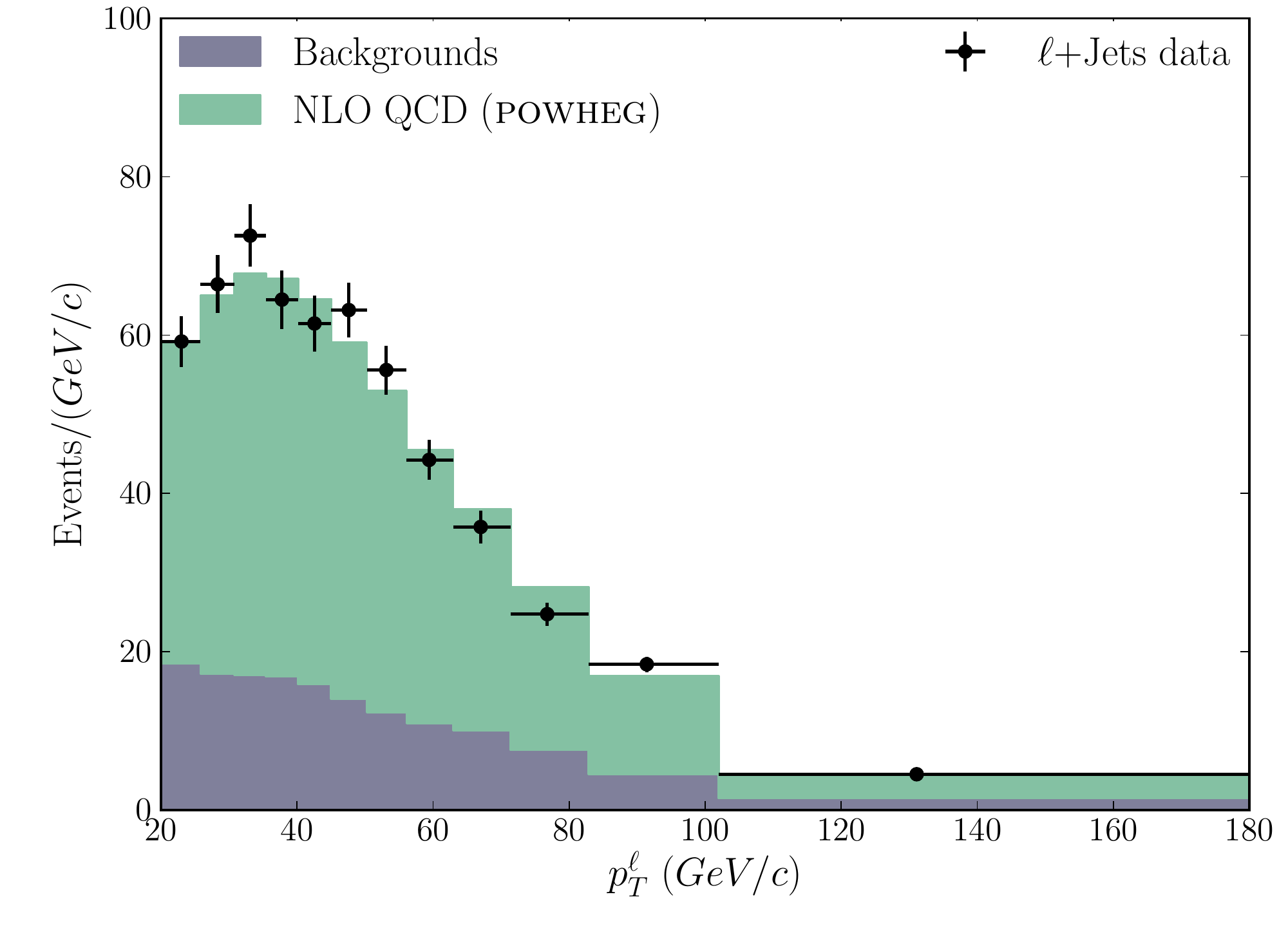}
\hspace{-0.2in}
\caption{{\small The distribution of candidate events in the signal region vs. the measured $p_{T}$ of the lepton.} \label{fig:TagPtLep}}
\end{figure}

\begin{figure*}[!h]
  \subfloat[][\label{fig:DataYq}]{
    \includegraphics[clip]{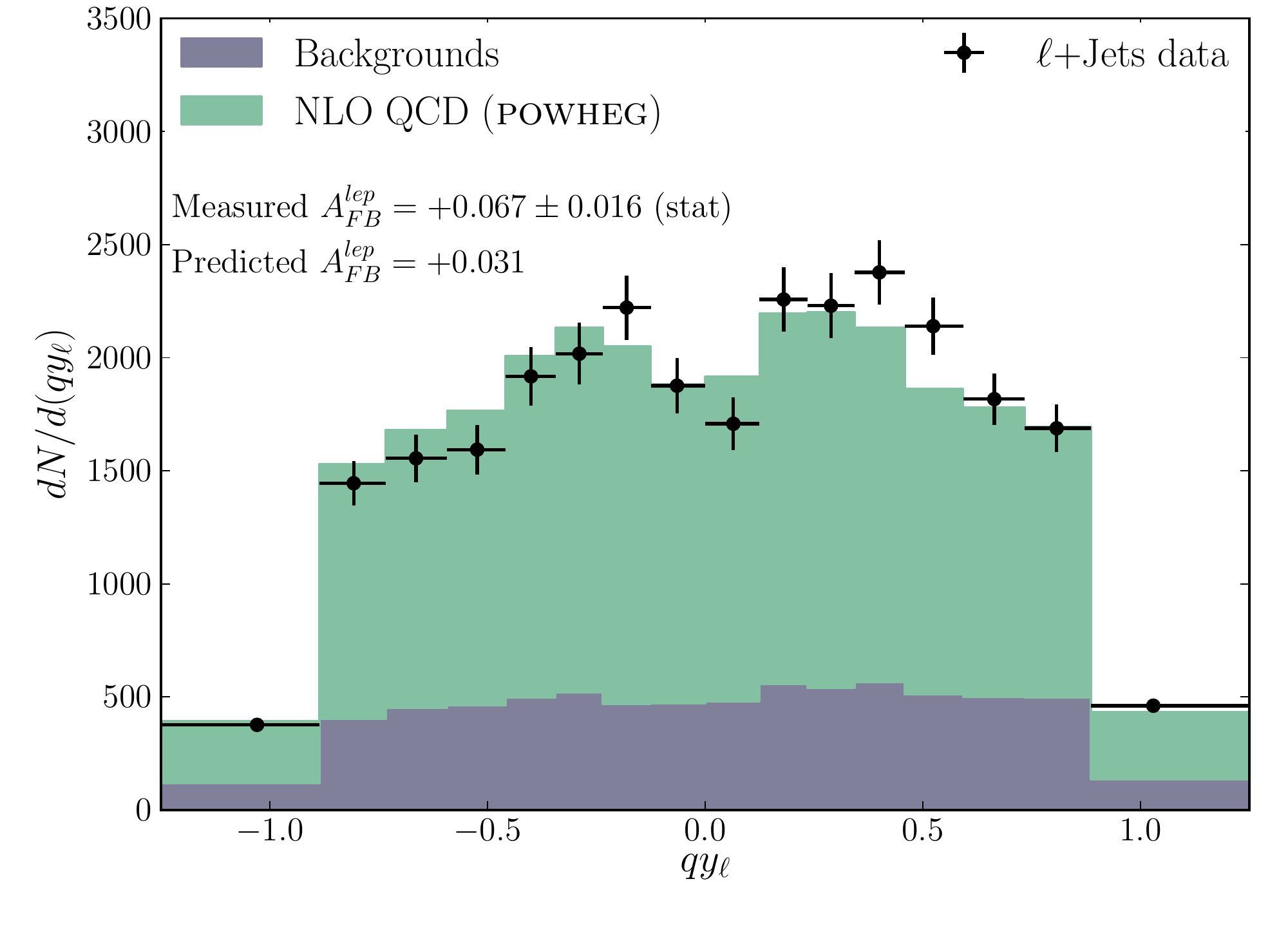}
  }
  \subfloat[][\label{fig:SigYq}]{
    \includegraphics[clip]{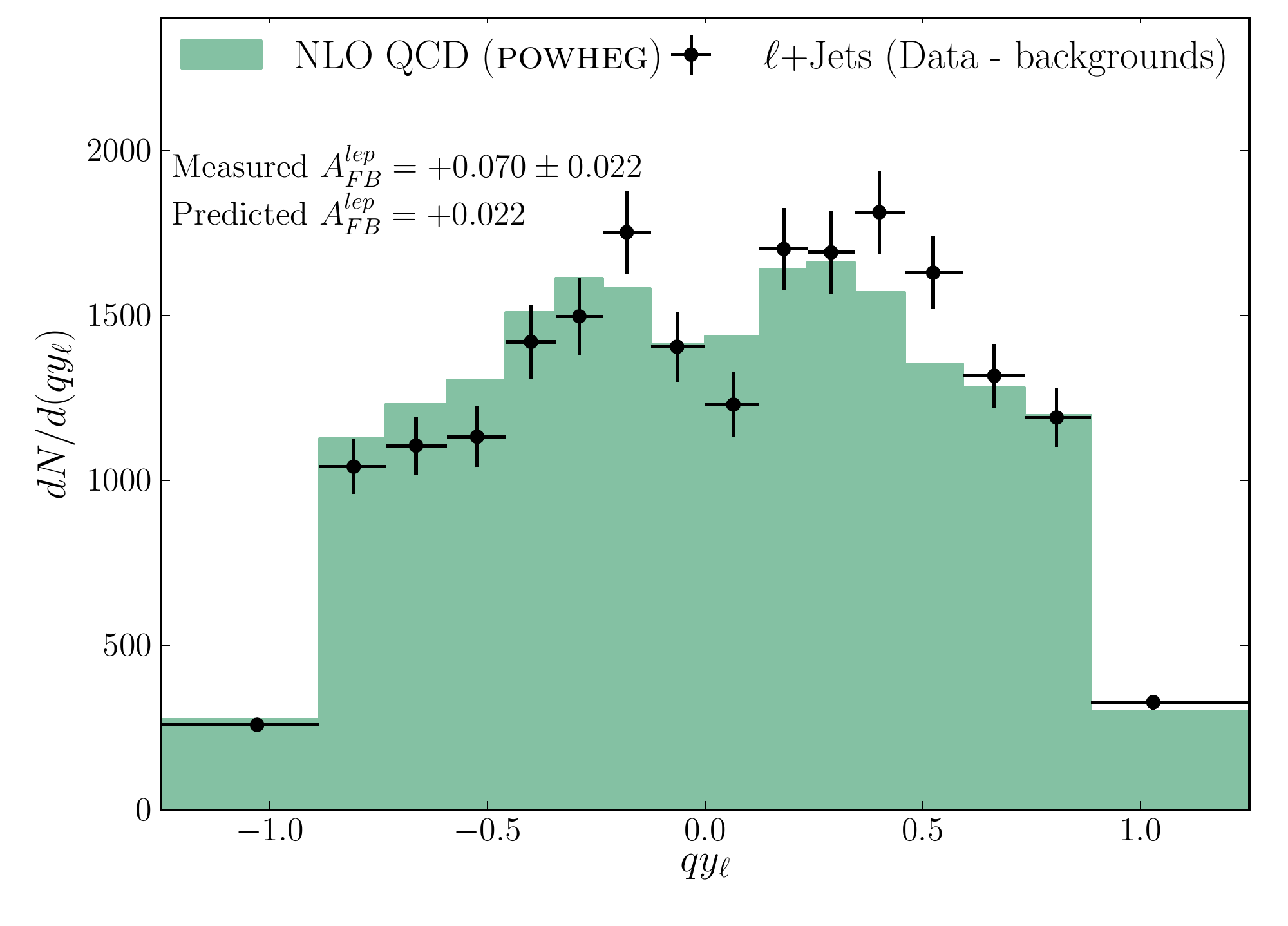}
  }
\caption{The observed distribution of events vs. $\yq$ in the signal region
\protect\subref*{fig:DataYq} compared to the NLO QCD prediction of {\sc powheg} and backgrounds; \protect\subref*{fig:SigYq}  after subtracting backgrounds, compared to the NLO QCD prediction of {\sc powheg}. 
\label{fig:Yq}}
\end{figure*}

We next examine the data during each stage of the analysis as outlined in Sec.~\ref{MethodValidation}.  We report values of $\afblep$ at several levels of correction: The {\it raw} $\afblep$ represents the complete and uncorrected selection; the {\it background-subtracted} asymmetry corresponds to a pure $\ttbar$, sample but it is not corrected for detector acceptance; and the {\it fully extrapolated} asymmetry is corrected to the production-level.  Unless otherwise noted, reported errors include both the statistical uncertainty as well as the systematic uncertainties appropriate to that correction level.

The modeling of the CDF $\ell$+jets data set has been extensively discussed and validated in Ref.~\cite{Aaltonen:2012it}. For the purpose of this analysis, we reproduce one associated distribution of interest --- the $\ptran$ of the lepton, shown in Fig.~\ref{fig:TagPtLep}. The {\sc powheg} signal model, along with non-$\ttbar$ background models and their normalizations, are seen to provide an accurate representation of the data.

The observed event distribution vs. the measured $\yq$ is shown in Fig.~\ref{fig:DataYq}.  The inclusive asymmetry observed in the data is $0.067\pm0.016$, compared to the predicted value of $0.031$ from {\sc powheg} and backgrounds.  Figure~\ref{fig:SigYq} shows the distribution of $\yq$ after backgrounds are subtracted.  The inclusive asymmetry is $0.070\pm0.022$.

The background-subtracted $\yq$ distribution is next decomposed into the corresponding $\Spartf$ (Fig.~\ref{fig:SigSpart}) and $\Apartf$ (Fig.~\ref{fig:SigApart}) parts.  The distribution of $\Spartf$ is in good agreement with the {\sc powheg} expectation.  The measured $\Apartf$ exceeds the predicted value in most bins, but becomes negative near $\left|\yq\right|=0$.  As the distribution of $\yq$ is expected to be continuous, its asymmetric part $\Apart$ must necessarily vanish as $\yq\rightarrow0$.  Consequentially, the observed deviation from this behavior must be statistical in nature.

\begin{figure*}
  \subfloat[][]{
    \includegraphics[clip]{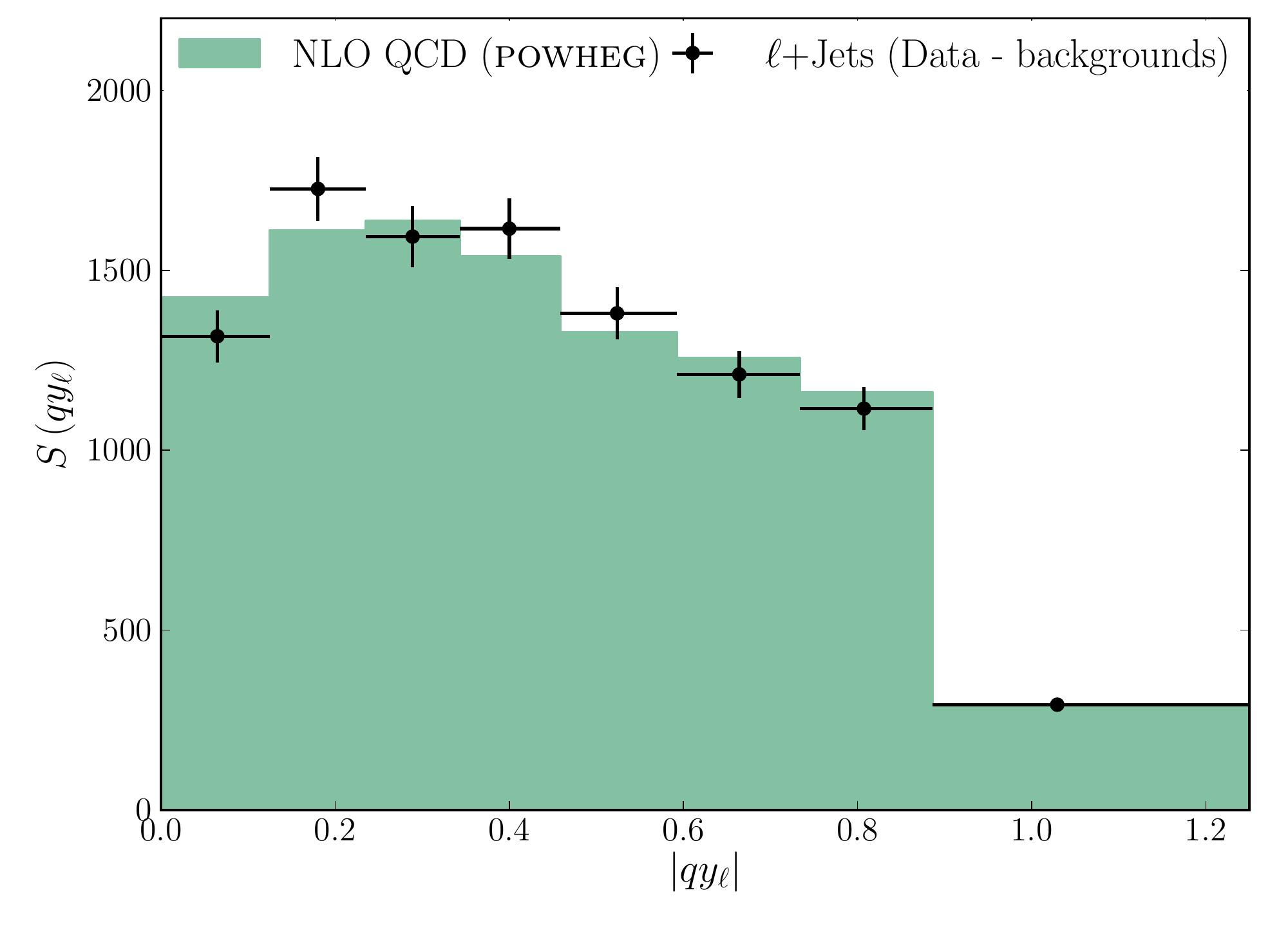}
    \label{fig:SigSpart}
  }
  \subfloat[][]{
    \includegraphics[clip]{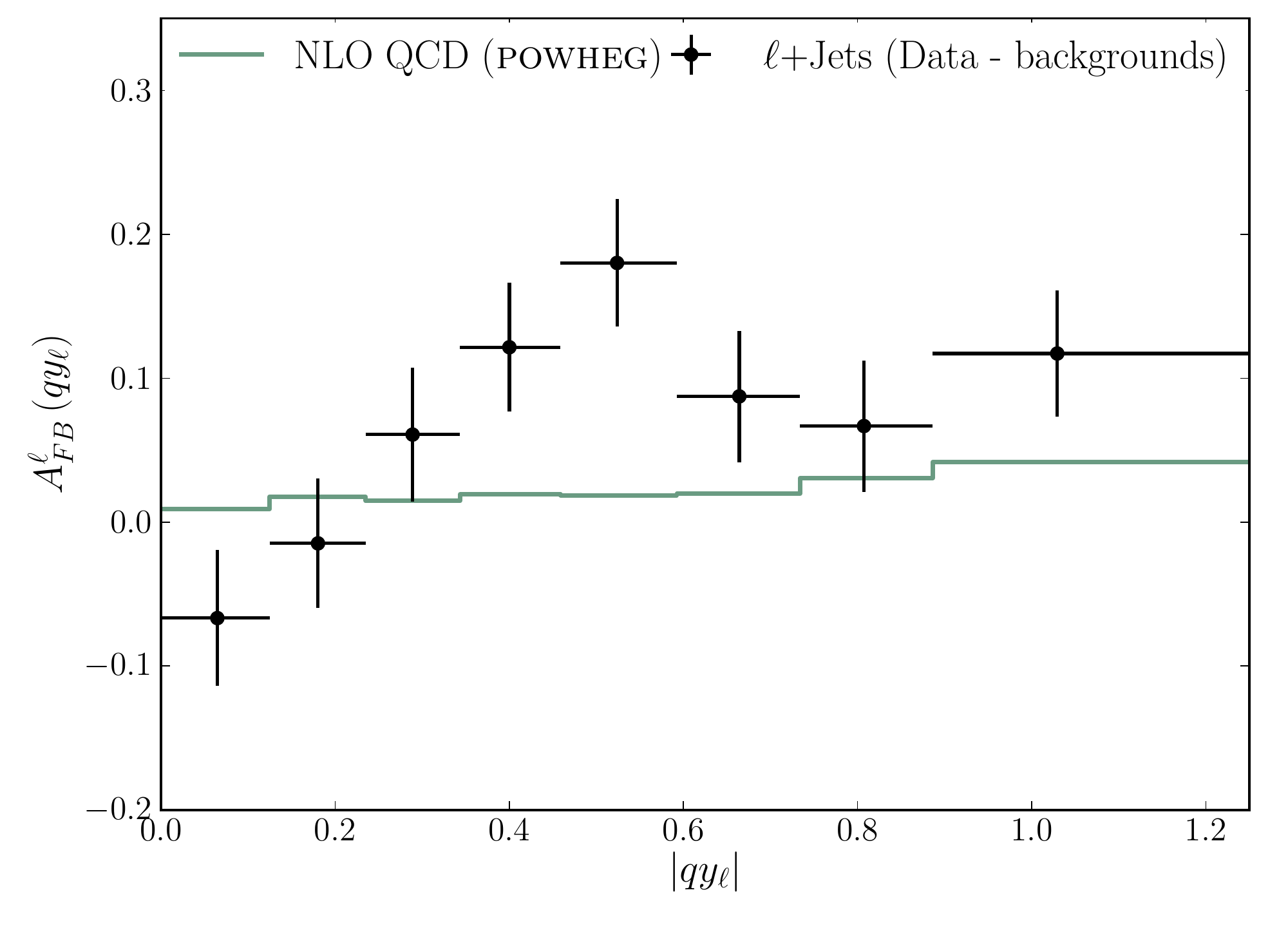}
    \label{fig:SigApart}
  }
\caption{ The symmetric part \protect\subref*{fig:SigSpart} and asymmetry \protect\subref*{fig:SigApart} as function of $\yq$ resulting from the decomposition of Fig.~\ref{fig:SigYq}. Data are shown as black markers, compared to the light-colored NLO QCD prediction of {\sc powheg}. \label{fig:YqDataNoFits}}
\end{figure*}

Acceptance corrections are then applied to the background-subtracted $\Apartf$ value, and the result is fit to Eq.~(\ref{eqn:APartFit}).  The acceptance-corrected data, {\sc powheg} prediction, and fits to both are shown in Fig.~\ref{fig:DataAFits}.  The estimated value of $a$ in the data is $0.266\pm0.079$ (stat.).  After performing the integration,  the resulting inclusive asymmetry in the data is $\afblep = 0.094\pm 0.024$.  This uncertainty is statistical only and is taken from the variance of the {\sc powheg} pseudoexperiments of Sec.~\ref{MethodValidation}.

\begin{figure}
\includegraphics[clip]{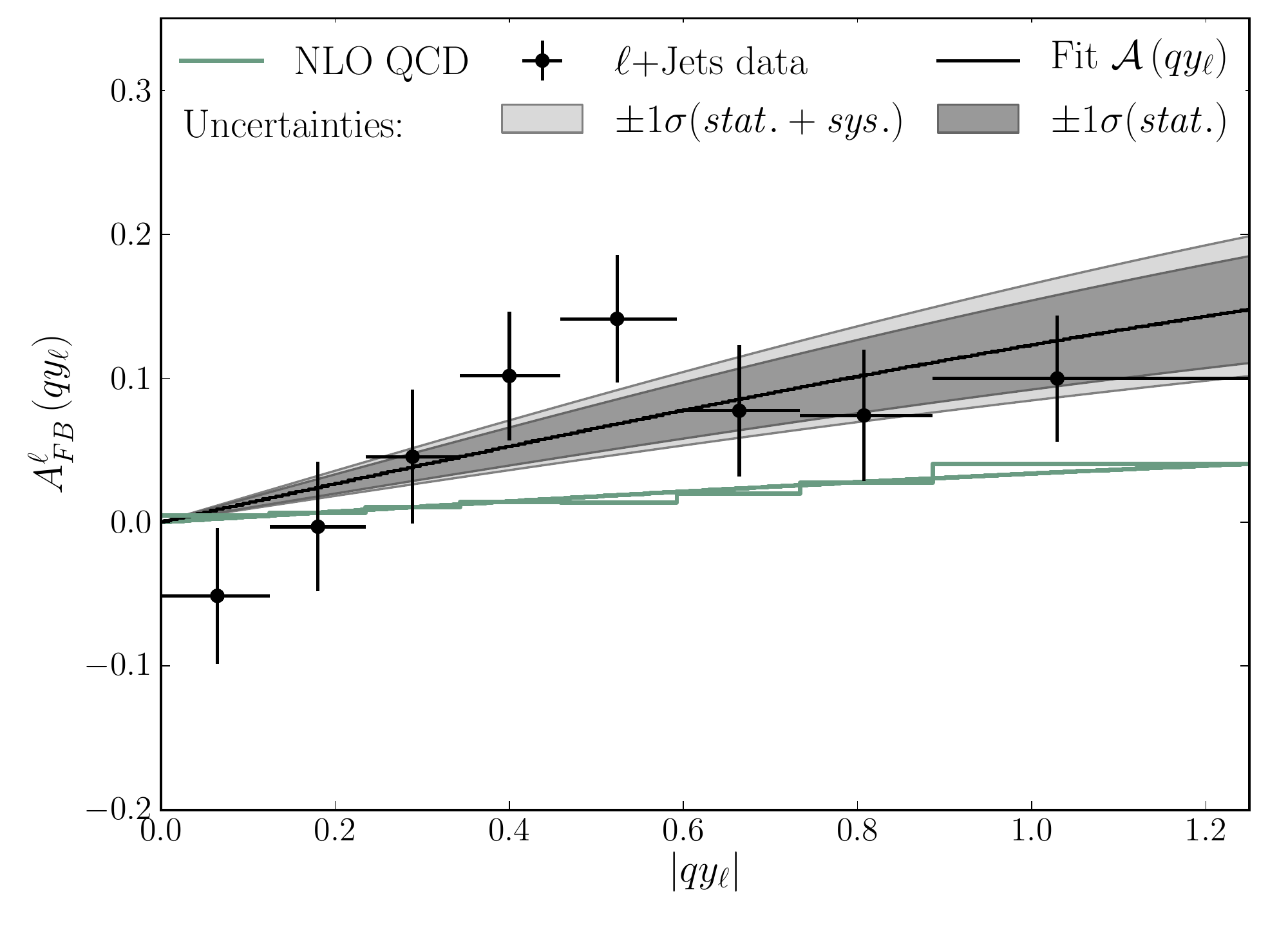}
\hspace{-0.2in}
\caption{ The binned asymmetry $\Apartf$ after correcting for acceptance, compared to the NLO QCD prediction of {\sc powheg}.  The best fit to Eq.~(\ref{eqn:APartFit}) for each is shown as the smooth curve of the same color.  The dark (light) gray bands indicate the statistical (total) uncertainty on the fit curve to the data.} \label{fig:DataAFits}
\end{figure}

\subsection{Evaluation of Uncertainties\label{InclAsymErr}}

The largest systematic uncertainty is associated with the background subtraction, where it is assumed that each background component has precisely the normalization reported in Table~\ref{tab:method2} and the statistically asymptotic shape of its prediction. The effects of uncertain normalizations and finite bin population are accommodated by extending the pseudoexperiment technique of Sec.~\ref{MethodValidation}. For each simulated experiment, a normalization for each signal and background component is randomly generated from a Gaussian distribution, using the expected event count and uncertainty.  Then the event count of each bin of each normalized component is randomly varied according to Poisson statistics. A set of $10~000$ simulated experiments is generated using {\sc powheg} as the signal model and subject to the entirety of the correction procedure.  This simultaneously incorporates the effects of statistical fluctuations on the bin populations and background shapes as well as the uncertainties on the expected background normalizations.

Another large uncertainty stems from the modeling of the $\ttbar$ recoil from QCD radiation.  The presence of radiated jets is strongly correlated with both $\afb$ and the $p_T$ of the $\ttbar$ system~\cite{Abazov:2011rq,Aaltonen:2012it,Skands:2012mm,Winter:2013xka}.  Color predominantly flows from an initiating light quark to an outgoing top quark (and from $\bar q$ to $\bar t$).  Events in which this color flow changes abruptly must radiate in order that the overall color current be conserved.  Consequentially, events in which the directions of the initiating light quark and outgoing top quark ($\bar{q}$ and $\bar{t}$) are different are typically associated with more radiation than those in which they are similar --- backward events ($\dy<0$) tend to radiate more than forward events ($\dy>0$).  The resulting larger average $p_T^{t\bar{t}}$ of backward events promotes them into the analysis sample with greater probability, inducing a small backward-favoring asymmetry in the acceptance of the lepton.

We assess an uncertainty on the modeling of this effect by comparing the result using the nominal {\sc powheg} model to other models.  We find that the recoil spectra of {\sc pythia} and {\sc alpgen} showered with {\sc pythia} are harder than {\sc powheg} showered with {\sc pythia}, resulting in larger acceptance corrections, increasing $\afblep$ by $0.013$.  We include a one-sided systematic uncertainty to reflect the fact that models other than {\sc powheg} are likely to increase the measured value of the asymmetry. An additional recoil-related bias may arise from the initial-and final- state radiation model (IFSR) of the {\sc pythia} showering of {\sc powheg}. We test this by studying the effect of reasonable variations in the amount of IFSR and find that the effect is small. 

\begin{table}
\caption{Uncertainties on the fully-extrapolated measurement.}\label{tab:ErrorTable}
\begin{tabular}{lr@{.}l}
\hline
\hline
Source of uncertainty & \multicolumn{2}{c}{Value} \\
\hline
                       Backgrounds & $0$ & $015$\\[0.7ex]
  \multirow{2}{*}{Recoil modeling} & $+0$ &$013$\\
                                   & $-0$ &$000$\\[0.7ex]
                Color reconnection & $0$ & $0067$\\
                  Parton showering & $0$ & $0027$\\
     Parton distribution functions & $0$ & $0025$\\
                 Jet-energy scales & $0$ & $0022$\\
 Initial and final state radiation & $0$ & $0018$\\
\hline
 \multirow{2}{*}{Total systematic} & $+0$ & $022$\\
                                   & $-0$ & $017$\\[0.7ex]
                  Data sample size & $0$ & $024$\\
\hline
\multirow{2}{*}{Total uncertainty} & $+0$ & $032$ \\
                                   & $-0$ & $029$ \\
\hline
\hline
\end{tabular}

\end{table}

Uncertainties on the signal model, including the above, enter only through the bin-by-bin acceptance corrections.  This class of uncertainties is quantified by performing the correction procedure on the data using acceptances from alternate simulated $\ttbar$ samples.  We also test the effects of color reconnection, parton showering, and jet-energy-scale (JES) uncertainties, all of which are small, as expected since jets are used only to define the signal region.  Uncertainties on the PDFs also have minimal impact. 

Table~\ref{tab:ErrorTable} summarizes all of the uncertainties considered. The largest uncertainty is due to the limited sample size. Combining the systematic uncertainties in quadrature we obtain the final result $\afblep = 0.094\pm 0.024^{+0.022}_{-0.017}$.

\subsection{Consistency Checks\label{InclSummary}}
\begin{table*}
\caption{Summary of asymmetries observed in subsamples selected by charge, lepton type, and jet multiplicity.  Exclusive categories are grouped together by horizontal lines.  Also reported is the inclusive result.  Uncertainties include both statistical and systematic contributions.}\label{tab:CrossChecks}
\begin{tabular}{lllcc}
\hline
\hline
Sample & Event yield & Raw & Background-subtracted & Fully extrapolated \\
\hline
   Electrons    & $1788$ & $0.050 \pm 0.024$ &  $0.050 \pm 0.033$ &  $0.062^{+0.052}_{-0.049}$\\
   Muons        & $2076$ & $0.081 \pm 0.022$ &  $0.087 \pm 0.029$ &  $0.119^{+0.039}_{-0.037}$\\
\hline
   Positive     & $1884$ & $0.099 \pm 0.023$ &  $0.110 \pm 0.031$ &  $0.125^{+0.043}_{-0.041}$\\
   Negative     & $1980$ & $0.036 \pm 0.022$ &  $0.034 \pm 0.031$ &  $0.063^{+0.046}_{-0.042}$\\
\hline
  $W$+$4$       & $2682$ & $0.064 \pm 0.019$ &  $0.064 \pm 0.024$ &  $0.084^{+0.035}_{-0.032}$\\
  $W$+$3$+$1$   & $1182$ & $0.072 \pm 0.029$ &  $0.092 \pm 0.049$ &  $0.115^{+0.067}_{-0.065}$\\

\hline
   Inclusive    & $3864$ & $0.067 \pm 0.016$ &  $0.070 \pm 0.022$ &  $0.094^{+0.032}_{-0.029}$\\
\hline
\hline
\end{tabular}

\end{table*}

To further check the validity of the inclusive measurement of $\afblep$, we divide the sample into several subsamples, which are expected to have the same inclusive asymmetries, summarized in Table~\ref{tab:CrossChecks}.

Two independent subsamples are formed by partitioning according to lepton flavor.  The raw asymmetry for decays into muons is $ 0.081 \pm 0.022 $ while that for decays into electrons is $ 0.050 \pm 0.024 $.  The difference is consistent with zero at about the $1\sigma$ level.  This difference is carried through each stage of correction  with similar levels of significance at each, resulting finally in fully-corrected asymmetries of $0.119^{+0.039}_{-0.037}$ in events with a muon and $0.062^{+0.052}_{-0.049}$ in events with an electron.

The sample is also partitioned according to lepton charge.  The difference between the raw asymmetries of the two subsamples is nonzero at $2\sigma$.  A similar difference is observed in the background-subtracted asymmetries. This difference is due to unphysical negative-asymmetry bins in the negatively-charged leptons near $\left|\yq\right|=0$. The fit, which by construction has $\mathcal{A}\left(0\right)=0$, is insensitive to these bins.  This moderates the discrepancy in the extrapolated result to $1\sigma$ after the extrapolation procedure is performed.

Finally, the sample is partitioned according to the $\etran$ of the fourth jet.  The first subsample consists of events having a fourth jet with $\etran > 20$~\gev .  This is the ``$W$+$4$'' jet selection used in Ref.~\cite{Aaltonen:2012it}.  In the present work we also include events with a $W$ and three jets with $\etran > 20$~\gev, isolating the $\ttbar$ component by requiring the presence of a fourth soft jet with $20\ge\etran>12~\gev$. This ``$W$+$3$+$1$'' sample shows consistent asymmetries with the $W$+$4$ sample at all levels of correction.

\section{Conclusions\label{Conclusions}}

The rapidity distribution of the lepton in semileptonic top quark decays contains information on the top-quark-production asymmetry and possible top-quark polarization, and is free of the complications of reconstruction the kinematic properties of the full $\ttbar$ system.  We develop a technique to measure the production-level lepton asymmetry in $\ell$+jets events, including an extrapolation to unmeasured rapidity regions, and apply it in a sample of 3864 $\ttbar$ candidate events collected with the CDF II detector at the Fermilab Tevatron. The production-level lepton asymmetry is found to be $\afblep = 0.094^{+0.032}_{-0.029}$. This is consistent with a value $\afblep = 0.111\pm 0.036$ measured by the D0 collaboration~\cite{Abazov:2012bfa}.  The present result is to be compared with the predicted value of $0.038\pm0.003$~\cite{Bernreuther:2012sx}, which includes both QCD and electroweak effects to NLO.  For a $\dy$ asymmetry as indicated by the Tevatron measurements, the expected lepton asymmetry is estimated to lie in the range $0.070$--$0.076$. 

\begin{acknowledgments}
We acknowledge the kind assistance of A. Falkowski and T. Tait in the construction of the Octet models as well as W. Bernreuther and G. Perez for helpful discussion.

We thank the Fermilab staff and the technical staffs of the participating
institutions for their vital contributions. This work was supported by the U.S.
Department of Energy and National Science Foundation; the Italian Istituto
Nazionale di Fisica Nucleare; the Ministry of Education, Culture, Sports,
Science and Technology of Japan; the Natural Sciences and Engineering Research
Council of Canada; the National Science Council of the Republic of China; the
Swiss National Science Foundation; the A.P. Sloan Foundation; the
Bundesministerium f\"ur Bildung und Forschung, Germany; the Korean World Class
University Program, the National Research Foundation of Korea; the Science and
Technology Facilities Council and the Royal Society, UK;  the Russian
Foundation for Basic Research; the Ministerio de Ciencia e Innovaci\'{o}n, and
Programa Consolider-Ingenio 2010, Spain; the Slovak R\&D Agency; the Academy of
Finland; the Australian Research Council (ARC); and the EU community Marie
Curie Fellowship contract 302103. 

\end{acknowledgments}

\bibliography{cites}

\begin{thebibliography}{27}
\bibitem [{coo()}]{coords}%
  \BibitemOpen
  \href@noop {} {}\bibinfo {note} {The polar angle is $\theta$ and the
  azimuthal angle is $\phi$. With total energy $E$ and momentum $p$, the
  transverse energy is defined as $E_{T} = E\sin\theta$ and the transverse
  momentum is $p_{T} = p\sin\theta$. The missing transverse energy (\met) is
  the magnitude of $\vec{\met} = -\Sigma_{i}E_{T}^{i}\hat{n}_i$ where
  $\hat{n}_i$ is a unit vector perpendicular to the beam axis and pointing to
  the $i$th calorimeter tower. The pseudorapidity is $\eta =
  -\ln(\tan(\theta/2))$.}\BibitemShut {Stop}%
\bibitem [{con()}]{conventions}%
  \BibitemOpen
  \href@noop {} {}\bibinfo {note} {$\Spart$ and $\Apart$ represent the
  continuous symmetric part and asymmetry. $\Apartf$ and $\Spartf$ are the
  binned equivalents of these continuous distributions. Simulated samples have
  very high statistics and may be binned finely enough to be treated as
  essentially continuous, and so are represented with $\Apart$ and
  $\Spart$.}\BibitemShut {Stop}%
\end{thebibliography}

\end{document}